\documentclass[12pt]{iopart}
\pdfoutput=1

\usepackage{iopams}  
\usepackage{color}
\usepackage {graphicx}
\def\>{\right\rangle}
\def\<{\left\langle}
\def\be{\begin{equation}}
\def\ee{\end{equation}}
\def\ba{\begin{array}{lll}}
\def\ea{\end{array}}

\def\beq{\begin{eqnarray}}
\def\eeq{\end{eqnarray}}
\begin{document}

\title{Finite frequency noise spectroscopy for fractional Hall states at $\nu=5/2$}

\author{A Braggio$^{1,2}$, M Carrega$^{1}$, D Ferraro$^{3}$ and M Sassetti$^{4,1}$}
\address{$^1$ SPIN-CNR, Via Dodecaneso 33, 16146 Genova, Italy}
\address{$^2$ I.N.F.N. Sezione di Genova, Via Dodecaneso 33, 16146 Genova, Italy}
\address{$^3$ D\'epartement de Physique Th\'eorique, Universit\'e de Gen\`eve, 24 quai Ernest Ansermet, CH-1211 Geneva, Switzerland}
\address{$^4$ Dipartimento di Fisica, Universit\`a di Genova, Via Dodecaneso 33, 16146 Genova, Italy}

\address{}
\ead{alessandro.braggio@spin.cnr.it}
\vspace{10pt}
\begin{indented}
\item[]\today
\end{indented}

\begin{abstract}
We investigate the finite frequency noise of a quantum point contact at filling factor $\nu=5/2$ using a weakly coupled resonant LC circuit as a detector. We show how one could spectroscopically address the fractional charged excitations inspecting separately their charge and scaling dimensions. We thus compare the behaviour of the Pfaffian and the anti-Pfaffian non-Abelian edge states models in order to give possible experimental signatures to identify the appropriate model for this fractional quantum Hall states. Finally we investigate how the temperature of the LC resonant circuit can be used in order to enhance the sensibility of the measurement scheme.  
\end{abstract}

\section{Introduction}
Among the extremely remarkable properties of the fractional quantum Hall (FQH) effect~\cite{Tsui99} a major role is played by the emergence of anyonic excitations carrying fractional charge and statistics~\cite{Stern08}. In particular, quasiparticle (qp) excitations for states belonging to the Laughlin \cite{Laughlin83} and Jain sequence~\cite{Jain89} are predicted to have Abelian exchange statistics. More intriguingly, some of the proposed models for the filling factor $\nu=5/2$~\cite{Willett87} predict the emergence of excitations with charge $e^{*}=e/4$, and multiples, with possible non-Abelian properties~\cite{Moore91}. These predictions paved the way to possible applications of non-Abelian anyons in fault-tolerant topological quantum computation (see~\cite{Nayak08} and references therein). Unfortunately, as far as we know, a direct confirmation of fractional statistics is still lacking even if different proposal are reported in the literature~\cite{Bishara09,Stern10,Willett10, Rosenow12}. So far, evidence of the fractional statistics is indirect, essentially based on the evidence of the existence of fractional charges~\cite{Laughlin83}. 

A great experimental effort has been devoted in the last years to access fractional charges through shot noise measurements in quantum point contact (QPC) geometry starting with the seminal works of Refs.~\cite{dePicciotto97, Saminadayar97}.
In this direction,  composite states (for example $\nu=2/5, 2/3$) showed a quite universal phenomenology leading to a crossover between two different value of the effective charge as a function of temperature or bias~\cite{Chung03, Bid09, Dolev10}. This behaviour has been explained in terms of competition of two charge carriers: the agglomerates and the single-qp~\cite{Ferraro08, Ferraro10, Ferraro10b}. Similar arguments hold as well for the state at $\nu=5/2$~\cite{Carrega11}. Other interpretations based on edge state reconstruction \cite{Wang13}, local filling factor effects~\cite{Roddaro04,Roddaro05,Hashisaka15} or tunnelling amplitude non-linearities~\cite{Shtanko14, Smits14} have also been proposed. Unfortunately, in the discussed measurements, the contributions associated to the various excitations are typically mixed because these studies are conducted at very low frequencies (almost DC). Therefore, it is useful to find alternative methods to address the excitations separately.

A possible way is to consider the noise at finite frequency (f.f.)~\cite{Rogovin74}. Indeed, this quantity presents resonant singular behaviour (such as peaks or dips) in correspondence of the Josephson frequency $\omega_{q}= q V/\hbar$ associated to each charge carrier $q$ with $V$ the applied bias. This is an independent method to measure the charge of the fractional excitations in the system that has not yet be experimentally explored for FQH states so far. Indeed, for sufficiently low temperatures $k_B T\ll qV$, it allows to separate the different charge contributions realising a sort of qp spectroscopy.
 Intriguingly enough, f.f. noise could efficiently  address also other properties, like the scaling dimensions associated to each qp excitation using only a bias scan at fixed frequency as we will show.
 
This combination of information has the potential to give further constraints on the edge state model~\cite{Radu08, Feldman14}, and finally to address the topological order of the bulk ground state~\cite{Susskind95}.  First theoretical steps in this direction were done on symmetrised noise~\cite{Chamon95, Chamon96}, typical quantity considered at low frequencies. In such case the expected features associated to the state at $\nu=5/2$~\cite{Bena06} as well as the possibility to access the contribution associated to the different qps predicted by theoretical models~\cite{Carrega12} have been investigated. However, at high frequencies $\omega  \gtrsim k_B T$, quantum effects become relevant and the symmetrised f.f. noise is only one possible choice among different  experimental quantities addressed by different protocols. Indeed, in such regime, one has to identify the relevant quantity measured with the specific setup under investigation \cite{Bednorz13}.
 Hereafter we get inspiration by the proposal of Lesovik and Loosen~\cite{Lesovik97} where a model based on a resonant 
LC circuit is discussed in order to extract non-symmetrised current-current correlations. Recent experiments carried out for a two dimensional electron gas QPC in absence of magnetic field fully agree with theoretical predictions~\cite{Zakka07}. Since the resonant circuit coupled to a QPC is the prototypical measurement scheme in FQH experiments~\cite{dePicciotto97, Saminadayar97} it appears quite obvious to explore the same physics in this contest.  We have recently investigated the same setup for an Abelian FQH states~\cite{Ferraro14}, and here we consider it to the case $\nu=5/2$ analysing the signatures of the different non-Abelian phases (Pfaffian and anti-Pfaffian). The goal of the present paper is to analyse in details the expected f.f. measured noise for a realistic situation. The effects associated to the temperature of the FQH QPC system and the LC detector on the visibility are also carefully considered.

The paper is divided as follows: In section 2 we discuss the non-Abelian models for edge states pointing out similarities and differences in term of the most dominant fractionally charged excitations. In section 3 we discuss the definition of the noise properties for a QPC in the weak-backscattering regime and the definition of the noise power measured in the proposed setup. In section 4 we discuss the result for the measured noise power obtained for two considered non-Abelian models and we also compare them with the well know symmetrised noise. Finally we inspect the effect of changing the detector temperature $T_c$.  

\section{Model}
We start recalling the two more accredited models of composite edge states
at filling factor $\nu=5/2$~\cite{Willett87}: the Pfaffian (P)~\cite{Moore91, Fendley07} and the disorder dominated anti-Pfaffian (AP)~\cite{ Lee07,Levin07}.  The associated Lagrangian densities are given by the sum of charged ($\mathcal{L}_{\mathrm{c}}$) and neutral contributions($\mathcal{L}_{\mathrm{n}}$), 
namely $\mathcal{L}=\mathcal{L}_{\mathrm{c}}+\mathcal{L}_{\mathrm{n}}$, 
with  ($\hbar=1$) 
\be \mathcal{L}_{\rm{c}}=-\frac{1}{2\pi}
\partial_{x}
\varphi_{\rm{c}}\left(\partial_{t}\varphi_{\rm{c}}+v_{\rm{c}}
  \partial_{x} \varphi_{\rm{c}}\right)
\label{lagrangian_charged}
\ee and 
\be \mathcal{L}_{\rm{n}}=-i\psi \left(\xi\partial_{t}
  \psi+v_{\rm{n}} \partial_{x}\psi\right)-\frac{\alpha}{4 \pi}
\partial_{x} \varphi_{\rm{n}} \left(\xi\partial_{t} \varphi_{\rm{n}}
  +v_{\rm{n}} \partial_{x}\varphi_{\mathrm{n}}\right).
\label{lagrangian_neutral}
\ee 
They both describe an Hall fluid at filling factor $\nu=1/2$ with two additional filled Landau levels, playing the role of the \emph{vacuum} of
the theory, according to the conventional decomposition $5/2=1/2+2$. The charged bosonic field $\varphi_{\rm{c}}(x)$ is related
to the electron number density through $\rho(x)=\partial_{x}\varphi_{\mathrm{c}}(x)/2 \pi$, while
$\varphi_{\rm{n}}(x)$ is a bosonic neutral field and $\psi(x)$ represents a
neutral Majorana fermion in the Ising sector~\cite{Stern10}.
  The parameter $\xi = \pm
1$ denotes the direction of propagation of neutral modes with respect
to the charged ones. In particular, for $\xi=+1$ the modes are co-propagating, while for $\xi=-1$ they are counter-propagating. 

The two models differ in the neutral sector $\mathcal{L}_{\mathrm{n}}$~\cite{Boyarsky09}, with $\alpha=0$ and $\xi=1$ for P, $\alpha=1$ and $\xi=-1$ for AP.
The propagation velocities of the charged and neutral modes are
indicated with $v_{\mathrm{c}}$ and $v_{\mathrm{n}}$ respectively. Due to the hidden symmetry of the neutral sector of AP model~\cite{Levin07} in the disorder dominated phase, one has the same velocity $v_{\mathrm{n}}$ both for the bosonic $\varphi_{{\rm n}}$ and the fermionic neutral modes $\psi$. Moreover one may reasonably assume a larger charge velocity $v_{\mathrm{c}}\gg v_{\mathrm{n}}$~\cite{Dolev10, Carrega11, Carrega12, Hu09}. The quantization of the above bosonic fields is given by the commutation relation  
\be 
\left[\varphi_{\rm{c/n}}(x),
\varphi_{\rm{c/n}}(y)\right]=i\pi \nu_{\rm{c/n}} {\rm{sgn}}(x-y),
\ee
with $\nu_{\rm{c}}=1/2$ and $\nu_{\rm{n}}=\xi $, while the Majorana fermion commutes with both.

\subsection{Quasiparticle operators and scaling dimensions}

Operators destroying an excitation along the edge can be written as \cite{Stern08,Fendley07, Levin07} 
\begin{eqnarray}
\Psi_{\rm P}^{(\chi,m)}(x)&\propto& \chi(x)
e^{i\frac{m}{2}\varphi_{\mathrm{c}}(x)}\nonumber\\
\Psi_{\rm AP}^{(\chi, m,
  n)}(x)&\propto& \chi(x)
e^{i\left[\frac{m}{2}\varphi_{\mathrm{c}}(x)+\frac{n}{2}\varphi_{\mathrm{n}}(x)\right]}\,,
\label{Psi}
\end{eqnarray}
with $m,n$ integer numbers and where the operator in the Ising sector $\chi(x)$ can be the identity operator $I$,
the Majorana fermion $\psi(x)$ or the spin operator $\sigma(x)$. They are associated  to the fact that the excitation charge  is even or odd multiple of the fundamental charge of the model $e^*=(e/4)$. Indeed all the excitations described by previous operators have charge $(m/4)e$ and we call them $m$-agglomerates~\cite{Ferraro08, Carrega11}.  The single-valuedness properties of the phase acquired by an
$m$-agglomerate with respect to the operation of encircling an electron in the bulk\footnote{The holographic principle~\cite{Susskind95} impose that the same restriction applies to the edge theory.} , force $m$ and $n$ to be: even integers for $\chi=I$ or $\psi$, and odd integers for $\chi=\sigma$~\cite{Stern10, Bishara08}. Notice that the presence of $\sigma$ in the operator leads to non-Abelian statistical properties important for fault-tolerant quantum computation as determined by the fusion rules~\cite{Stern08, Nayak08}.

The zero temperature time dependent Green's functions associated to the operators of the Ising sector and the
charged and neutral bosonic fields
are~\cite{Ferraro10, Ferraro10b, Ginsparg89, Weiss99, Cuniberti96, Braggio00, Braggio01} 
\beq
\langle \chi(0,t) \chi(0,0) \rangle&=(1+i \omega_{\mathrm{n}}t)^{-\delta_{\chi}},\qquad &\label{GFIsing}\\
\langle \varphi_{s}(0,t) \varphi_{s}(0,0) \rangle&=-|\nu_{s}| \ln{
  (1+i \omega_{s} t)} \qquad &s=\mathrm{c},
\mathrm{n}\label{GFbosonic} \eeq
with $\delta_{I}=0$, $\delta_{\psi}=1$ and $\delta_{\sigma}=1/8$ the conformal weights of the field in the Ising sector~\cite{Stern08, Nayak08, Ginsparg89}
and we introduced the energy bandwidths
$\omega_{\mathrm{c/n}}=a^{-1}v_{\mathrm{c/n}}$, with $a$ a finite
length cut-off.  In the following we will assume $\omega_{\rm c}$ as the
largest energy scale of the model. From the long-time behaviour of the imaginary time
two-point Green's function
\footnote{Notice that, for notational convenience, we omitted any reference to the neutral sectors in the upper index in operator of the vertex $\Psi^{(m)}_{l}$ and in the scaling $\Delta^{(m)}_{l}$. From now on indeed we will consider for any $m$-agglomerate only the qp operators with the minimal scaling dimension compatible with the single-valuedness requirement.}  
\be
\langle T_{\tau}\Psi^{(m)}_{l}(\tau) {\Psi_{l}^{(m)}}^\dagger(0)\rangle
\propto |\tau|^{-2\Delta^{(m)}_{l}}\qquad l=\mathrm{P, AP}
\ee
we can extract the scaling dimensions~\cite{Kane92} of the $m$-agglomerates
\be \Delta_{\rm
  P}^{(m)}=\frac{1}{2}\delta_{\chi}+\frac{1}{16} m^{2};\qquad
\Delta_{\rm AP}^{(m)}=\frac{1}{2}\delta_{\chi}+\frac{1}{16}
m^{2}+\frac{1}{8}n^{2}
\label{Delta}
\ee
which depends on the model considered.
Therefore, for the single-qp with minimal charge $e^{*}=e/4$ ($m=1$, $\chi=\sigma$ and only for AP $n=\pm 1$) one has respectively
\be
\Delta_{\rm P}^{(1)}=\frac{1}{8};\qquad
\Delta_{\rm AP}^{(1)}=\frac{1}{4},
\label{1_8}
\ee
while the $2$-agglomerate excitation with charge $e/2$ ($m=2$, $n=0$, $\chi=I$), with a scaling dimension driven by the charged mode contribution only with 
\be 
\Delta_{\rm P}^{(2)}=\Delta_{\rm AP}^{(2)}=\frac{1}{4}.
\label{deltaagg}
\ee 
These values indicate the single-qp as the most dominant excitation at low energy in the P case, while in the 
AP case single-qp and 2-agglomerate have equal relevance with the same scaling dimensions~\cite{Levin07}. The latter situation is quite general and valid for all anti-Read-Rezayi states~\cite{Braggio12a}.  All other excitations, with higher charges, have higher scaling dimensions and can be safely neglected in what follows.
It is worth to mention that interactions with the external environment can lead to renormalizations of the scaling parameters with remarkable consequences on the transport properties (see Ref.~\cite{Braggio12} for a better discussion). In the following, for sake of simplicity, we will focus on the unrenormalised case only despite the method  may be generalised to the renormalised case.

\section{Noise properties in QPC at finite frequency} 
\begin{figure}[ht]
\centering
\includegraphics[scale=0.7]{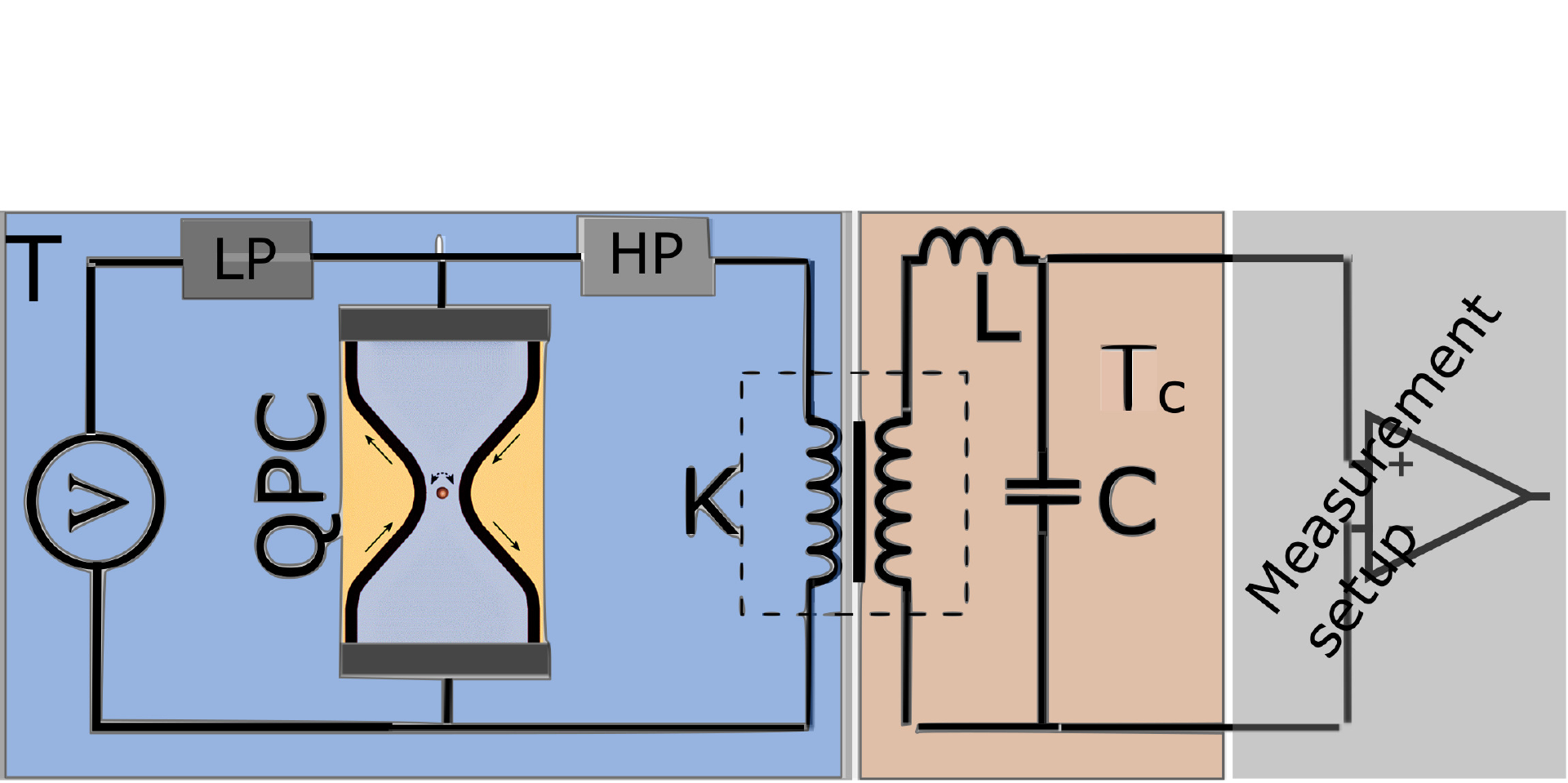}
\caption{Schematic view of an Hall bar with a QPC (blue area) weakly coupled with LC detector (yellow area). A bias voltage $V$ is applied to the QPC and $m$-agglomerate excitations can tunnel between the edges. The two circuits are matched in impedance  via a coupling circuit (inside dashed line). Here we assume they are kept at two different temperatures $T$ and $T_c$ as indicated in the picture. In the figure we explicitly show also with grey boxes a low-pass (LP) and high-pass (HP) filters which make possible to directly couple the DC bias to QPC and, at the same time, to deviate the high-frequency components toward the LC circuit.}
\label{fig1}
\end{figure}

Once characterised the excitations of the considered models for $\nu=5/2$, we can investigate the associated f.f. backscattering noise in the QPC geometry shown in Fig.~\ref{fig1}. A similar measurement scheme was proposed for the first time by Lesovik and Loosen in Ref.~\cite{Lesovik97}. Here, the QPC is subjected to a bias voltage V and coupled to a resonant LC circuit, playing the role of the detector (with measurement frequency $\omega=\sqrt{1/LC}$), via an impedance matching circuit (see dashed box in Fig.~\ref{fig1}). With strong magnetic field the impedance matching in the system is a challenging technological problem~\cite{Altimiras13, Altimiras14} therefore it is advantageous to suppose to work at fixed resonant frequency $\omega$ assuming a very high quality factor of the detector. 

We focus on the simple two terminal geometry in the weak backscattering limit, where $m$-agglomerate tunnelling processes can be treated separately. In real systems a four terminal version of this setup is required, however this doesn't change the main result obtained with this simplified version. 

The point-like tunnelling of a generic $m$-agglomerate between the right-propagating ($+$) and the left-propagating ($-$) edge can be described through the tunnelling Hamiltonian ($l=\textrm{P,AP}$)
\be
\label{tun}
{\cal H}^{(m)}_{T, l}= t_{m}\  \Psi_{l, +}^{(m)}(0){\Psi^{(m)}_{l, -}}^{\dagger}(0)+h.c.,
\ee
where $t_{m}$ is the  $m$-agglomerate tunnelling amplitude (assumed energy independent). The finite bias $V$ between the two edges can be included in our formalism through the gauge transformation $t_{m}\to t_{m}e^{i m \omega_0t}$,  where, $\omega_0=e^*V/\hbar$ is the Josephson frequency 
associated to the fundamental charge $e^*$~\cite{Martin05}. From the tunnelling Hamiltonian in~(\ref{tun}) one can easily derive the backscattering current operator associated to the $m$-agglomerate~ \cite{Ferraro08, Ferraro10, Ferraro10b, Carrega11}
\be
I^{(m)}_{B, l}(t)=i me^* \left(t_{m}\  e^{im\omega_0t}\Psi_{l,+}^{(m)}(0,t){\Psi^{(m)}_{l,-}}^{\dagger}(0,t)-h.c.\right).
\ee
The contribution to the averaged backscattering current of the $m$-agglomerate at lowest order in the tunnelling can be easily written in terms of the tunnelling rates $\mathbf{\Gamma}^{(m)}_l$~\cite{Carrega12}
 \be \langle I^{(m)}_{{B,l}} \rangle=m e^{*}\left(1-
  e^{-m\omega_0/T}\right)\mathbf{\Gamma}^{(m)}_l ( \omega_0)\,,
 \label{current_bal} \ee
 with the average $\langle...\rangle$ taken over the quantum statistical ensemble. The rates can be also evaluated analytically at low temperatures $T\ll\omega_{{\rm n}},\omega_{{\rm c}}$~\cite{Ferraro08,Carrega12}
\beq
\mathbf{\Gamma}^{(m)}_l (\omega_0) &=&\frac{|t_m|^2}{(2 \pi a)^2}\frac{ (2 \pi) ^{\eta_{m}+ \mu_{l, \alpha}}}{\omega_{\mathrm{c}}^{\eta_{m}} \omega_{\mathrm{n}}^{\mu_{l, \alpha}}}T^{\eta_{m}+\mu_{l, \alpha}-1}e^{\frac{m\omega_0}{2T}} \nonumber \\
 &\times& B\left(\eta_{m}+\mu_{l, \alpha}-i\frac{m\omega_0}{2 \pi T}; \eta_{m}+\mu_{l, \alpha}+i\frac{m\omega_0}{2 \pi T}\right)\
 \label{eq:Gamma}
\eeq
where $B(x,y)$ is the Euler beta function.
Here, $\eta_m=m^2/4$ and $\mu_{l, \alpha}=2\delta_{\chi} +\alpha n^2/2$ depend on the variables $m,\chi$ and $n$ which characterise the tunnelling excitation for the specific P or AP model considered (see discussion around~(\ref{1_8}) and~(\ref{deltaagg})).  Note that for $k_BT\ll \omega_0\ll\omega_{\rm n},\omega_{\rm c}$ asymptotic expansion shows that ${\bf\Gamma}^{(m)}_l (\omega_0)\propto |\omega_0|^{4 \Delta^{(m)}_{l}-1}$, with the expected power law dependences of the rates from the bias energy $\omega_0$ and the scaling dimension $\Delta^{(m)}_{l}$ as usually happen in the Luttinger liquid theory. For higher bias value $\omega_{\rm n}\ll\omega_0\ll\omega_{\rm c}$ the power-law does not depend anymore on the neutral components and one finds ${\bf\Gamma}^{(m)}_l (\omega_0)\propto |\omega_0|^{m^2/4-1}$ where the power-law scaling is determined only by the charge of the $m$-agglomerates. The low-energy analytical result presented corresponds to the standard golden rule rate for the tunnelling processes and one may eventually calculate it also with numerical methods following the prescription of Ref.~\cite{Carrega12}.

The proper quantity to consider in order to investigate the current fluctuations of the QPC coupled to the resonant circuit is the non-symmetrised noise~\cite{Lesovik97, Gavish00, Aguado00,Creux06,Zazunov07,Chevallier10}
\be
S^{(m)}_{+}(\omega)=\frac{1}{2}\int^{+\infty}_{-\infty} dt \, e^{i \omega t}
 \langle \delta I^{(m)}_{B,l}(0) \delta I^{(m)}_{B,l}(t) \rangle, 
\ee
where we have introduced the back-scattering current fluctuation $\delta I_{B,l}^{(m)}= I^{(m)}_{{B,l}}-\langle I^{(m)}_{B,l}\rangle$ \footnote{For notational convenience we omitted the index $l={\rm P},{\rm AP}$ on the noise power since its definition is exactly the same for the two models.}. 
This quantity represents, for $\omega>0$, the noise power emitted by the system into the detector. The corresponding absorptive part is given by 
\be
S^{(m)}_{-}(\omega)=\frac{1}{2}\int^{+\infty}_{-\infty} dt \, e^{i \omega t}
\langle \delta I^{(m)}_{B,l}(t) \delta I^{(m)}_{B,l}(0) \rangle=S_+^{(m)}(-\omega)
\label{S_minus}\ .
\ee
  With these quantities it is easy to calculate the f.f. symmetrised noise~\cite{Rogovin74, Chamon95, Chamon96,Carrega12,Blanter00} usually considered in literature
\beq
\label{eq:Ssym}
S^{(m)}_{sym}(\omega)&=& \frac{1}{2}\int^{+\infty}_{-\infty} d t \,e^{i \omega t}
\langle \{ \delta I^{(m)}_{B,l}(t), \delta I^{(m)}_{B,l}(0)\}\rangle
= S^{(m)}_{+}(\omega)+S^{(m)}_{-}(\omega)
\label{eq:s_sym}
\eeq
having indicated with $\{\cdot ,\cdot \}$ the anticommutator. At lowest order in the tunnelling amplitudes and using standard Keldysh formalism also the non-symmetrised noise can be expressed in terms of the QPC tunnelling rates~\cite{Ferraro14} 
 \be
S^{(m)}_{+}(\omega,\omega_0)= \frac{(m e^{*})^{2}}{2} \left[{\bf{\Gamma}}^{(m)}_l \left(-\omega+ m\omega_{0}\right)+{\bf{\Gamma}}^{(m)}_l \left(-\omega- m\omega_{0}\right)\right],
\label{S_plus_rate}
\ee
with a peculiar combination of the frequency $\omega$ and the bias voltage $\omega_0$ in the arguments of the golden rule rates.

The detector of Fig.~\ref{fig1} represents a concrete measurement scheme to access current fluctuations at high frequencies. In the following we will focus on the regime where the QPC temperature $T$ is lower than the frequency (quantum limit) and the bias (shot noise limit), i.e. $k_BT\ll e^*V,\omega$. This allow to investigate the fractional qp contributions via a sort of spectroscopy.

The measurable quantity is the spectral power measured in the amplifier chain (grey area in Fig.~\ref{fig1}), which is proportional to the variation of the energy stored in the LC before and after the switching on of the LC-QPC coupling. From now on we will indicate it as measured noise $S_{meas}(\omega,\omega_0)$ where with $\omega$ we indicate the frequency of the LC circuit and with $\omega_0$ the QPC bias. At lowest order in the coupling $K\ll 1$ it can be expressed as~\cite{Bednorz13,Lesovik97,Gavish00,Chevallier10,Gavish04, Gavish01} 
\be
S^{(m)}_{meas}(\omega,\omega_0)= K \left\{S^{(m)}_{+}(\omega,\omega_0)+n_{B}(\omega) \left[ S^{(m)}_{+}(\omega,\omega_0)-S^{(m)}_{-}(\omega,\omega_0)\right]\right\},
\label{S_meas}
\ee
where the non-symmetrised noise QPC spectra for the $m$-agglomerate are $S^{(m)}_{\pm}(\omega,\omega_0)$ of (\ref{S_plus_rate}).   
Here, $n_{B}(\omega)= 1/[e^{\beta_{c} \omega}-1]$
the Bose distribution describing the equilibrium state of the LC detector
and $\beta_{c}=1/k_{B}T_{c}$ the detector inverse temperature. In general $T_c$ can be different from the system temperature $T$ since system and detector are {\it weakly} coupled. We wish finally recall that this quantity can be also investigated using a strategy similar to the definition of the excess noise which further simplify the impedance matching problem at the level of the LC-QPC coupling (see Ref. \cite{Ferraro14} for details). 

To consider all contributions due to tunnelling of different $m$-agglomerate, in weak tunnelling, one can directly sum them~($ i=meas,sym$)
\be
S_{i}(\omega_0)= \sum_{m} S^{(m)}_{i}(\omega_0)\ ,
\label{noise_sum}
  \ee
where from now on we suppress the explicit dependence on the LC frequency $\omega$ since in the following discussion is always kept fixed. 
We conclude this part noting that these results suggest that f.f. noise is a spectroscopy tool for different tunnelling charges.

\section{Results and discussion}

\begin{figure}[ht]
\centering
\includegraphics[width=.49\linewidth]{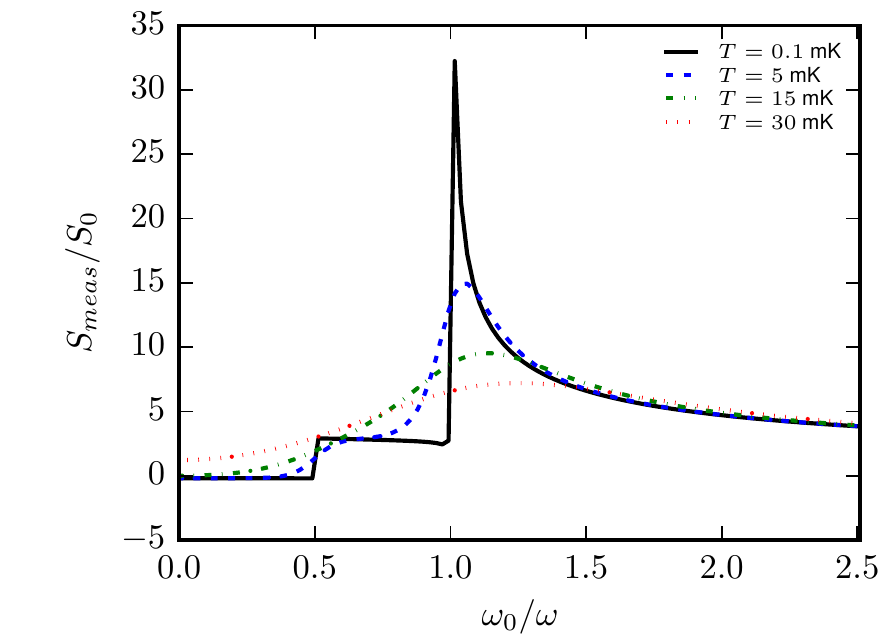}
\includegraphics[width=.49\linewidth]{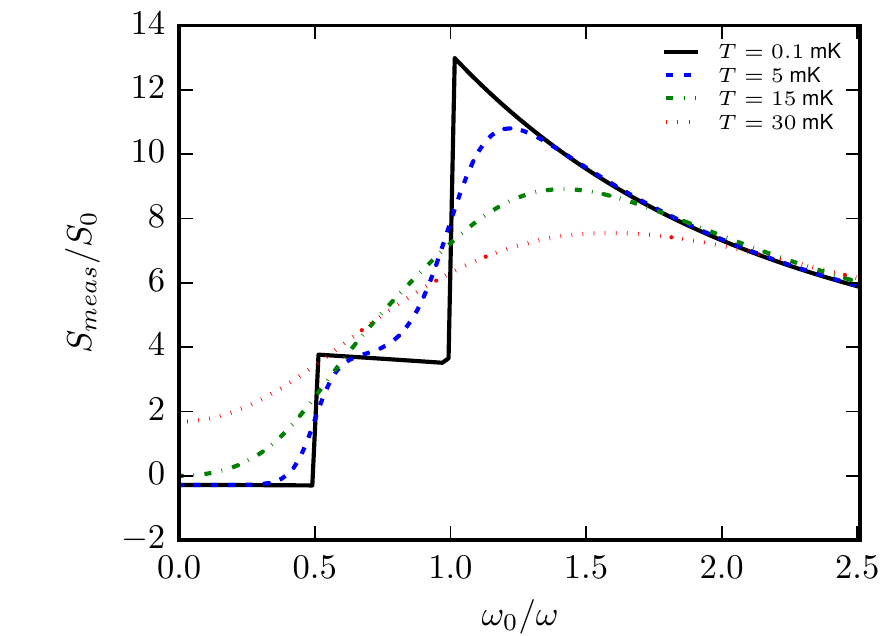}\\
\includegraphics[width=.49\linewidth]{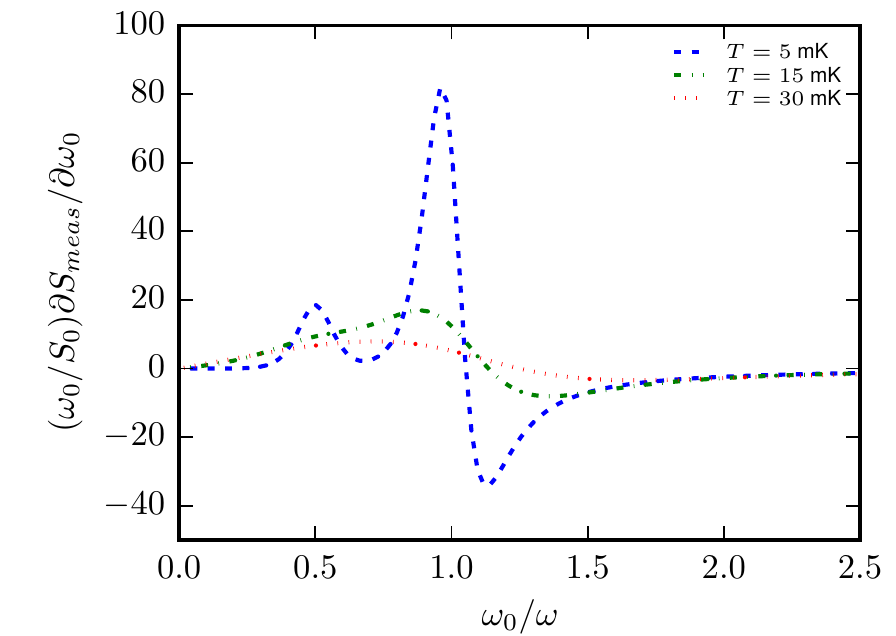}
\includegraphics[width=.49\linewidth]{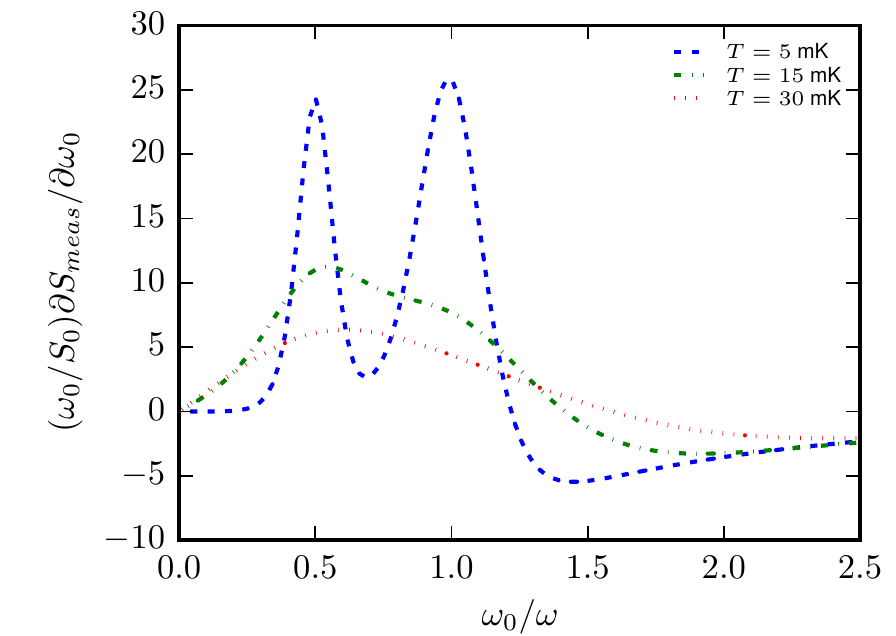}
\caption{(Upper panels) Measured f.f. noise $S_{meas}$ (in units of $S_0=K e^{2} |t_{1}|^{2}/(2\pi \alpha)^2\omega_{c}$) for the P (left) and the AP (right) model at $\nu=5/2$ as a function of the bias $\omega_{0}=e^*V$ and at fixed frequency $\omega$. The QPC temperature associated to each curve is indicated in the legend. (Bottom panels) Corresponding derivatives $\partial S_{meas}/\partial \omega_{0}$. The curve at the lowest temperature ($T=0.1$ mK) has been omitted for better visibility. Other parameters expressed in temperature scale are: $T_{c}=15$ mK, $\omega=60$ mK, $\omega_{n}=50$ mK, $\omega_{c}=1$ K and $|t_{2}|^2/|t_{1}|^2=1$.}
\label{fig2}
\end{figure}

The results concerning the measured f.f. noise $S_{meas}(\omega_0)$ in (\ref{S_meas}) and~(\ref{noise_sum}) at $\nu=5/2$ are shown in Fig.~\ref{fig2} (upper panels) as a function of the QPC bias $\omega_0$ for different QPC temperatures $T$, keeping fixed the resonant circuit temperature $T_c$. We discuss only the behaviour at positive bias $\omega_0>0$ since the noise is a symmetric function of the QPC bias $\omega_0$. 
Analogies and differences between the P (upper left panel) and AP (upper right panel) models become evident from a direct comparison. Starting from the lowest temperature case ($T=0.1$ mK, black curves) we observe that both models show a flat behaviour at $\omega_{0}/\omega\approx 0$, which is a clear signature of the lack of contribution of ground states fluctuations in the considered measurement scheme~\cite{Lesovik97}. The little deviation from zero are associated to the mismatch between the systems and detector temperature which can be always cancelled when $T=T_c$~\cite{Ferraro14}. Steep jumps associated to the $2$-agglomerate contribution appear at $|\omega_{0}|/\omega=1/2$ showing an identical profile in both models, which reflects the same scaling dimension of the two model for that excitation (see~(\ref{deltaagg})). Different is the spike associated with the single-qp occuring at $|\omega_{0}|/\omega=1$. Indeed, they are much more high and sharp in the P case with respect to the AP reflecting a lower scaling dimension of qp excitation for the P model. This feature could quite clearly distinguish between the two models. However the temperature should be kept quite  low, since increasing it the differences are progressively less marked. Eventually some signature survive only by considering the bias derivative of this quantity (see bottom panels of Fig.~\ref{fig2}).   

The previous behaviours can be explained in a simple way in the quantum limit ($k_{B}T_{c}\ll \omega$) for the detector and the shot-noise limit ($k_{B}T \ll \omega_{0}$) for the system. In this case one has the contributions of single and double excitations for the two models ($l=\textrm{P,AP}$)~\cite{Ferraro14}
\be
\label{shift}
S_{meas}(\omega_0)\approx \alpha_1 {\bf\Gamma}_l^{(1)}(\omega_0-\omega)+\alpha_2 {\bf\Gamma}_l^{(2)}(2\omega_0-\omega)
\ee
with $\alpha_1$ and $\alpha_2$ constant prefactors and the explicit expression  of the rates are reported in~(\ref{eq:Gamma}) for $\omega,\omega_0\ll \omega_{n},\omega_c$. 
This result confirms the same scaling for the $2$-agglomerate in the two models, but different behaviours for the single-qp contributions (see~(\ref{1_8}) and (\ref{deltaagg})). As we observed after (\ref{eq:Gamma}) for high biases $\omega_0\gg\omega_{\rm n}$ the scaling dimensions are determined only by the charged part that is the same for the two models. This is another reason why conventional scaling analysis, which is typically done in asymptotic regime, would fails in detecting the differences between the two models especially when the neutral mode bandwidth $\omega_{\rm n}$ is quite small\footnote{According to the condition $v_{\rm n}\ll v_{\rm c}$.}. 

By increasing the system temperature $T$, keeping fixed the one of the detector ($T_{c}=15$ mK), the peaked structure become progressively smoothened due to a rounding of the singularities. In this regime, the differences in the measured f.f. noise become clear only by looking the derivative with respect to the QPC bias $\partial S_{meas}/\partial \omega_{0}$  as shown in the bottom panels of Fig.~\ref{fig2}.  By focusing on the blue curves, and keeping in mind the different scale in the ordinates between the two panels, one can observe again the similarity of peaks associated to the $2$-agglomerate ($|\omega_{0}|/\omega=1/2$). However concerning the single-qp ($|\omega_{0}|/\omega=1$), the difference in the scaling leads to a pronounced peak followed by a stronger dip in the P case with respect to the AP case. 
\begin{figure}[ht]
\centering
\includegraphics[width=.49\linewidth]{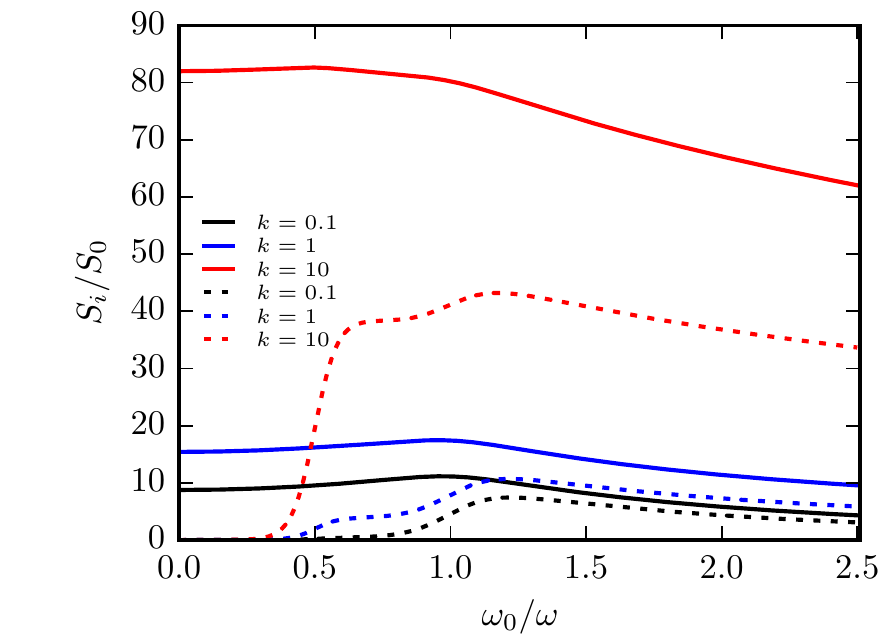}
\includegraphics[width=.49\linewidth]{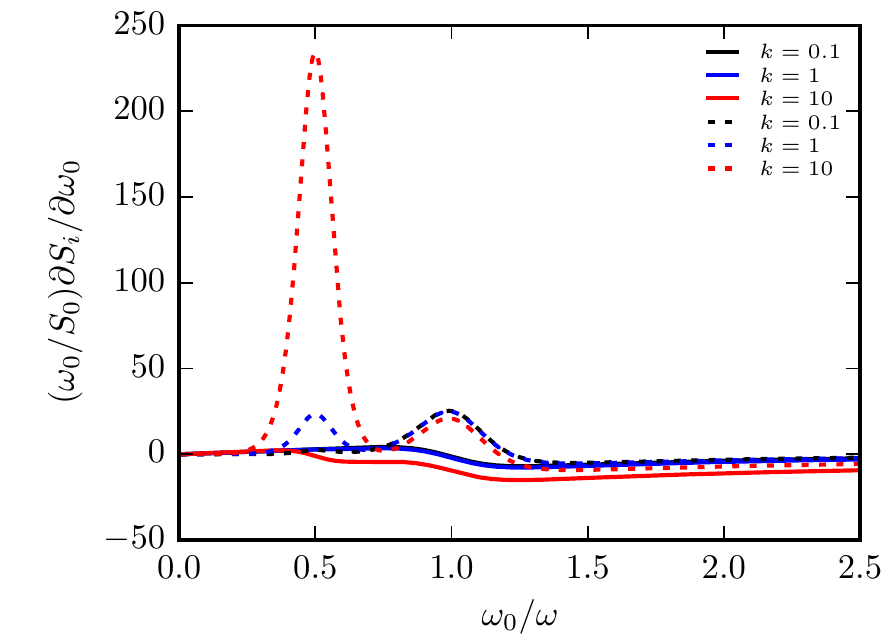}
\caption{(Left panel) Comparison for the AP model between the measured f.f. noise $S_{meas}$ (dashed lines) and the symmetrised noise and $S_{sym}$ (solid lines) for different values of the ratio $k=|t_{2}|^2/|t_{1}|^2$ (indicated in the legend).  (Left panel) The f.f. noises 
$S_{i}/S_{0}$ with $i=meas,sym$ in units of $S_0=K e^{2} |t_{1}|^{2}/(2\pi \alpha)^2\omega_{c}$. Note that $K=1$ for the symmetrised noise. (Right panel) The bias derivatives 
$\partial S_{i}/\partial \omega_{0}$ in units $S_0/\omega_0$. Other parameters are: $T=5$ mK, $T_{c}=5$ mK, $\omega_{n}=50$ mK and $\omega_{c}=1$ K.
}
\label{fig3}
\end{figure}

Until now all the plots were done for fixed ratio $k=|t_{2}|^2/|t_{1}|^2=1$. However this parameter is unknown and 
may change for any specific experimental realization.
For this reason we present in Fig.~\ref{fig3} with dashed lines $S_{meas}(\omega_0)$ (right panel) and its bias derivative  $\partial S_{meas}/\partial\omega_0$ (left panel) as a function of bias with changing $k$ values. We concentrate mainly on the AP model but similar considerations can be repeated in the P case.
 As expected, increasing this parameter progressively enhances the $2$-agglomerate contribution with respect to the single-qp but still leave both visible at different bias values. This is particularly true looking the bias derivative. This result shows the convenience of proposed setup in order to address the presence of the two different charged excitations also when eventually one of the contribution is deeply suppressed in comparison of the other due to non-universal effects. 

\begin{figure}[ht]
\centering
\includegraphics[width=.49\linewidth]{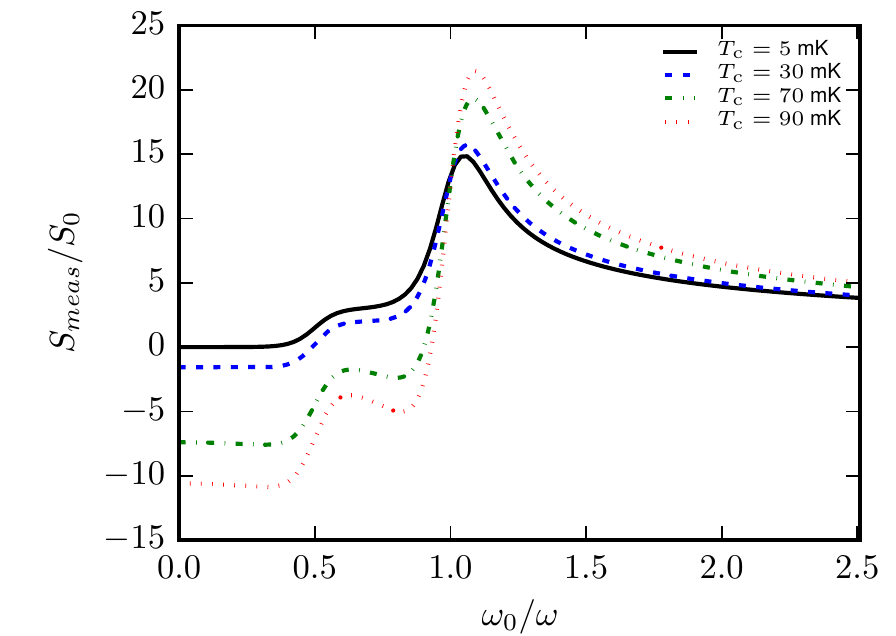}
\includegraphics[width=.49\linewidth]{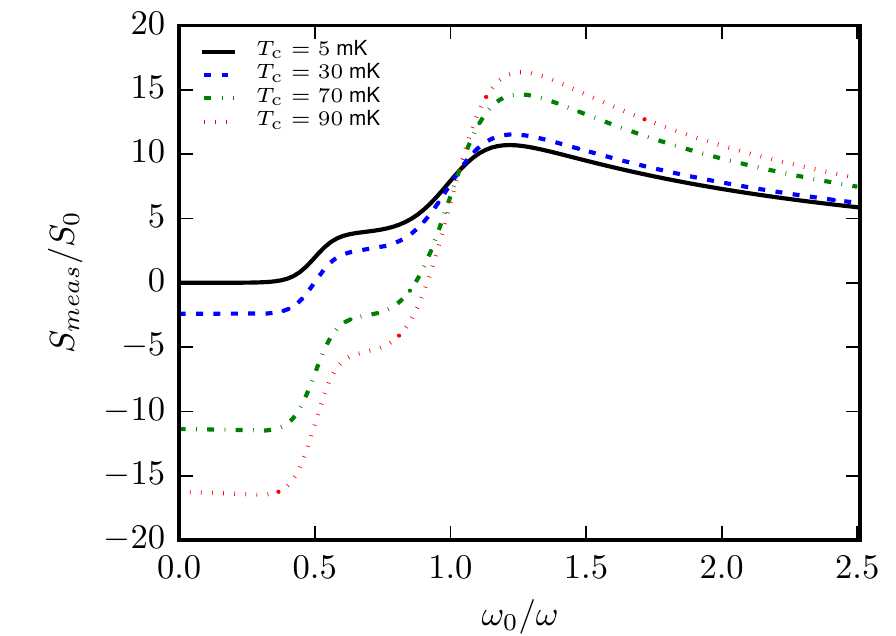}\\
\includegraphics[width=.49\linewidth]{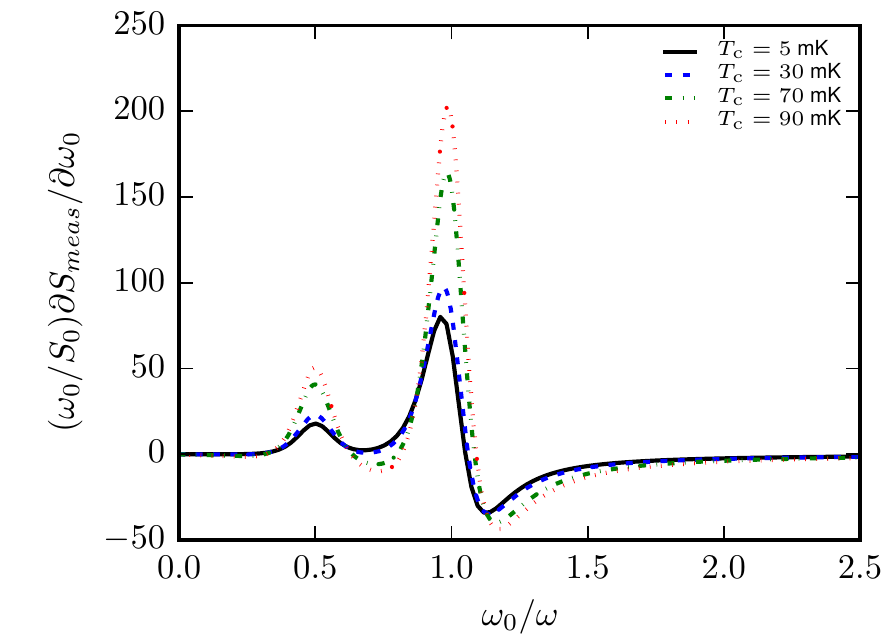}
\includegraphics[width=.49\linewidth]{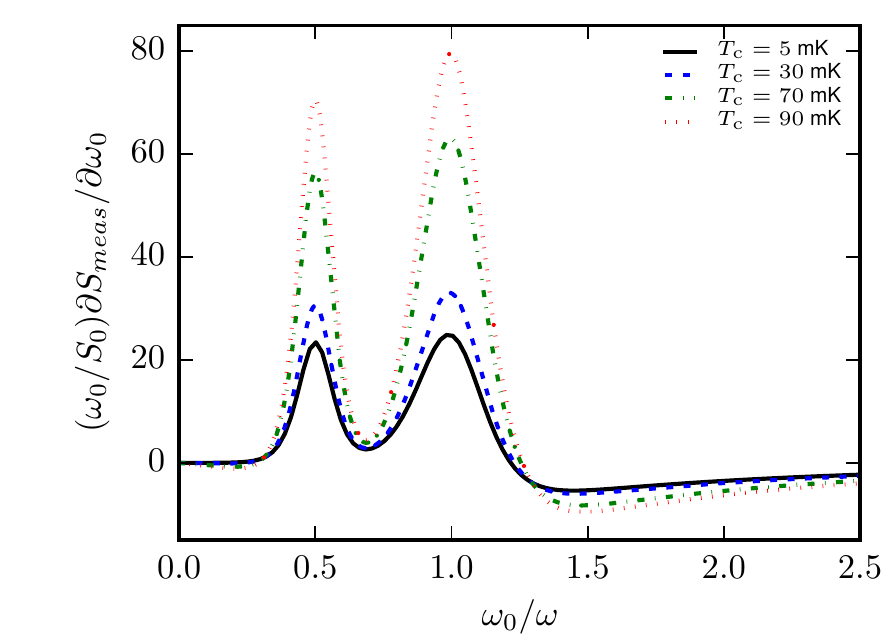}
\caption{(Upper panels) Measured f.f. noise $S_{meas}$ (in units of $S_0=K e^{2} |t_{1}|^{2}/(2\pi \alpha)^2\omega_{c}$) for the P (left) and the AP (right) model for the Hall state at $\nu=5/2$ as a function of the voltage and at fixed frequency ($\omega_{0}/\omega$) varying the detector temperature (see legend). (Bottom panels) Corresponding derivatives $\partial S_{meas}/\partial \omega_{0}$. Other parameters are (in temperature units where necessary): $T=5$ mK, $\omega=60$ mK, $\omega_{n}=50$ mK, $\omega_{c}=1$ K and $|t_{2}|^2/|t_{1}|^2=1$.}
\label{fig4}
\end{figure}

In order to make this statement more quantitative, in Fig.~\ref{fig3},  we compare $S_{meas}(\omega_0)$ with the f.f. symmetrised noise $S_{sym}(\omega_0)$ (solid lines) in~(\ref{eq:Ssym}) and their bias derivatives (in right panel)~\cite{Carrega12}.  
Here, again, we keep fixed the frequency $\omega$, changing the bias $\omega_0$ that is, by far, the most convenient protocol at the investigated GHz range\footnote{The impedance matching condition is much easier to be obtained at fixed frequency.}. We see that the $S_{sym}(\omega_0)$, as a function of the bias, is unable to detect the two singularities associated to the two fractional charges even in the bias derivatives. In particular, changing the tunnelling amplitude ratio $k$, the quantity seems only affected with a common multiplicative factor demonstrating that the signature of the two excitations is mainly mixed in that quantity. This supports the idea that in this bias dependent protocol $S_{sym}(\omega_0)$ is not useful especially in comparison to $S_{meas}(\omega_0)$. This statement can be easily verified by looking at both the left and right panel of Fig.~\ref{fig3}.

Finally, in order to find a signature of the charged excitations without identifying their scaling dimension, we could vary the LC detector temperature $T_c$ in order to increase the sensibility for charge detection~\cite{Ferraro14}. In Fig.~\ref{fig4} we show that this approach works for both the two non-Abelian edge models. Increasing $T_c$ increases the height of the jump in $S_{meas}(\omega_0)$ (top panels) which correspond also to an increase of the height of the peak in the derivatives (bottom).  Note that since the coupling with the detector is assumed weak (no poisoning from the detector) the width of the peaks is only slightly influenced by the detector temperature, preserving the resolving power of the discussed bias spectroscopy. Then the crucial limiting factor to the bias spectroscopy is the QPC temperature $T$.

As shown in Fig.~\ref{fig4} the power to distinguish between the AP and P model is not essentially compromised by the detector temperature since all the feature characterising the model in terms of $S_{meas}(\omega_0)$ seems mutually amplified. Increasing $T_c$ is an interesting resource in order to increase detection efficiency, in this perspective.

\section{Conclusions}
We have investigated the behaviour of f.f. emitted power $S_{meas}(\omega_0)$ of a resonant LC circuit weakly coupled to a QPC built in a quantum Hall bar at filling factor $\nu=5/2$. We showed that the emitted power is represented in terms of the non-symmetrised noise components of the quantum Hall QPC weighted by the bosonic distribution of the resonant LC circuit.
We have inspected the different predictions of the Pfaffian and anti-Pfaffian non-Abelian edge states models for this quantity. 
We showed that this setup can detect and discriminate between the dominant and sub-dominant fractionally charged excitations looking at the bias dependence at fixed GHz frequencies. We also discussed the advantage to use this measurement protocol in comparison to the  f.f. symmetrised noise.
 Finally we demonstrated how the sensibility of the proposed setup can be increased varying the LC detector temperature $T_c$.
\section*{Acknowledgements}
We thank C. Glattli, W. Belzig and P. Solinas for useful discussions.
We acknowledge the support of the MIUR-FIRB2012 - Project HybridNanoDev (Grant  No.RBFR1236VV), EU FP7/2007-2013 under REA grant agreement no 630925 -- COHEAT, MIUR-FIRB2013 -- Project Coca (Grant
No.~RBFR1379UX), and the COST Action MP1209. A.B acknowledges support from STM 2015 from CNR and Victoria University of Wellington where partially work was done.

\end{document}